\documentclass[twocolumn,superscriptaddress,showpacs,aps,floatfix,10pt]{revtex4-1}
\usepackage{amsmath,amsfonts,amssymb,epsfig,dcolumn,bm,dsfont,graphics,latexsym,color,graphicx,multirow,ulem,xcolor}

\let\oldAA\AA

\renewcommand{\AA}{\text{\normalfont\oldAA}}
\begin{document}
\preprint{AIP/123-QED}
\title{Hydrostatic pressure control of the spin-orbit proximity effect, spin relaxation, and thermoelectricity in a phosphorene-WSe$_2$ heterostructure}
\author{Marko Milivojevi\'c}
\email{marko.milivojevic@savba.sk}
\email{milivojevic@rcub.bg.ac.rs}
\affiliation{Institute of Informatics, Slovak Academy of Sciences, 84507 Bratislava, Slovakia}
\affiliation {Faculty of Physics, University of Belgrade, 11001 Belgrade, Serbia}
\author{Marcin Kurpas}
\affiliation{Institute of Physics, University of Silesia in Katowice, 41‑500 Chorz\'ow, Poland}
\author{Maedeh Rassekh}
\affiliation{Institute of Physics, Pavol Jozef \v{S}af\'{a}rik University in Ko\v{s}ice, 04001 Ko\v{s}ice, Slovakia}
\author{Dominik Legut}
\affiliation{IT4Innovations, VSB-Technical University of Ostrava, 17. listopadu 2172/15, 708 00 Ostrava, Czech Republic}
\affiliation{Department of Condensed Matter Physics, Faculty of Mathematics and Physics, Charles University, Ke Karlovu 3, 121 16 Prague 2, Czech Republic}
\author{Martin Gmitra}
\email{martin.gmitra@upjs.sk}
\affiliation{Institute of Physics, Pavol Jozef \v{S}af\'{a}rik University in Ko\v{s}ice, 04001 Ko\v{s}ice, Slovakia}
\affiliation{Institute of Experimental Physics, Slovak Academy of Sciences, 04001 Ko\v{s}ice, Slovakia}
\
\begin{abstract}
Effective control of interlayer interactions is a key element in modifying the properties of van der Waals heterostructures and the next step toward their practical applications. 
Focusing on the phosphorene-WSe$_2$ heterostructure, we demonstrate, using first-principles calculations, proximity-induced amplification of the spin-orbit coupling in phosphorene by applying vertical pressure.
We simulate external pressure by changing the interlayer distance between bilayer constituents and show that it is possible to tune the spin-orbit field of phosphorene holes in a controllable way. By fitting effective electronic states of the proposed Hamiltonian to the first principles data, we reveal that the spin-orbit coupling in phosphorene hole bands is enhanced more than two times for experimentally accessible pressures up to 17 kbar.
Correspondingly, we find that the pressure-enhanced spin-orbit coupling boosts the Dyakonov-Perel spin relaxation mechanism, reducing the spin lifetime of phosphorene holes by factor 4. We further explore the role of the lateral shift on the spin-orbit field and reveal that the spin-orbit strength of phosphorene holes can be sizably modulated when strong pressure is applied. We also found that the thermopower is governed mainly by the phosphorene and pressure reduces the overall thermoelectric efficiency of the heterostructure.
\end{abstract}
\maketitle
\section{Introduction}

Van der Waals (vdW) heterostructures~\cite{GG13,NMC+16,LWD+16,JMH17,XIZ+20} represent a unique class of materials and devices created by stacking together individual layers of two-dimensional materials held together by weak van der Waals forces. Using different combinations of materials, each with distinct properties, it is possible for one material to inherit the properties of the other via the proximity effect~\cite{B05,SFK+21}.

In the field of spintronics~\cite{WAB+01,ZFS04,FME+04,J12,HKG+14}, focusing on the manipulation of an electron's spin, it has been shown that vdW heterostructures can be used to modify sizably the spin-orbit coupling strength~\cite{GF15,GF17,ZGF19,SMK+23,MK2023} and induce magnetism~\cite{ZGF+16,ZGF18,ZJF19,ZGF22,ZF22} into the material of interest using the proximity effect. Phosphorene~\cite{LYY+14,RCN14,LNZ+14,FGF+15, LDD+15,LKJ+17,ATK+17}, a single layer of black phosphorus, is a semiconducting material with a sizable gap~\cite{SMG+16,ZHC+17,FDT+19,HFM+23} and the potential to overtake graphene~\cite{HKG+14} in future spintronics devices, provided that a sizable spin-orbit coupling and/or magnetism can be induced in a system.

Quite recently we have shown that combining materials with different crystal structures provides an excellent platform for the engineering of exotic spin texture in a low symmetry~\cite{MGK+23,MGK++23} or symmetry-free~\cite{MGK+24} environment. Specifically, in a mixed-lattice heterostructure made of phosphorene and WSe$_2$, a material with giant spin-orbit coupling~\cite{ZCS11,KBG+15}, we have shown that the sizable spin-orbit coupling can be induced in phosphorene holes, with the strengths exceeding the values when phosphorene is subjected to the perpendicular electrical field~\cite{KGF16,Kurpas2018:JPD}. 
Furthermore, we have shown that by symmetrical and asymmetrical encapsulation of phosphorene~\cite{MGK++23} by WSe$_2$ monolayers, one can tune both the strength and the type of spin texture induced in phosphorene holes. 

In this paper, we investigate another knob for the fine-tuning of the spin-orbit proximity effect, namely, the hydrostatic pressure. It has been demonstrated to be a promising approach for the modification of interactions between the vdW heterostructure constituents, resulting in a series of novel phenomena predicted theoretically~\cite{RMT14,CFH+18,LZN20} and experimentally~\cite{YJL+18,YCP+19,FMT+21,SRV+21,ZGL+22,WCY+22,PCM+23,FMZ+21}. Also, it was shown that the spin-orbit proximity effect depends on the interlayer distance between the heterostructure constituents~\cite{DRK+19}, and is tunable with hydrostatic pressure. Here we study electronic properties of the zero twist angle phosphorene-WSe$_2$ heterostructure showing that the spin-orbit coupling increases more than two times when assuming pressures up to 17 kbar. We note that such pressures are nowadays achievable in experimental setups~\cite{FMZ+21}. 
We also show that the tunability of the spin-orbit field by the pressure can be traced to the decreased spin relaxation times for the phosphorene holes. This represents an experimentally accessible way~\cite{ATK+17,CLT+24} to trace the spin-orbit strength engineered by pressure indirectly.

It has been predicted that both phosphorene and WSe$_2$ possess excellent thermoelectric properties~\cite{kumar2015thermoelectric,li2020recent,cui2021prediction, kumari2021possible,guo2023thermoelectric}, making them promising candidates for advanced energy conversion applications. The effective handling of thermoelectric materials is crucial in converting temperature gradient into electrical energy and vice versa,  which is essential for sustainable and efficient energy solutions.
Optimal thermoelectric properties of materials are highly desirable for future energy harvesting and environmentally friendly reasons \cite{Petsagkourakis2018:STAM}, such as the development of portable, solid state, passively powered electronic systems.
Figure of merit $ZT$, a dimensionless thermal to electric conversion quantity, can be enhanced through quantum confinement effects~\cite{Hicks1993:PRB}. These effects have been demonstrated on several phosphorene nanostructures~\cite{Szczech2011:JMC,Mazzamuto2011:PRB}, including nanorings that show high thermopower performance in the order of $\sim 5000~{\rm \mu V/K}$~\cite{Borojeni2023:PRB}.
Among the wide applications of black phosphorus, the wide tunability of its band gap~\cite{Gomez2015:JPCL} makes it suitable for thermoelectric applications such as thermal energy scavenging near room temperature.
For naturally hole-doped bulk black phosphorus, the Seebeck coefficient at room temperature is about $335~{\rm \mu V/K}$~\cite{Flores2015:APL}, and it can be further enhanced by ion gating~\cite{Saito2016:NL}. This tunability makes black phosphorus an excellent material for fine-tuning thermoelectric properties to optimize performance under various conditions.
The anisotropic electrical and lattice thermal conductance properties of black phosphorus~\cite{Qin2015:PCCP,Zhang2017:SR,Xu2018:MSMSE,Lee2019:SR} further enhance its suitability for thermoelectric applications. These properties result in pronounced directional dependencies, leading to orthogonal conducting directions~\cite{Fei2014:NL} which can be utilized to maximize thermoelectric efficiency.
Studies on strain effects have revealed that both black phosphorus and phosphorene are potential materials for medium- to high-temperature thermoelectric applications~\cite{Qin2014:SR,Lv2014:PRB}.
Furthermore, in semiconducting nanostructures, the breakdown of the Wiedemann-Franz law allows for the possibility of exceeding the conventional upper limit of $ZT$~\cite{Dubi2011}, thereby boosting the potential of vdW heterostructures for practical thermoelectric applications.
Motivated by these findings, in the final section, we will also explore thermoelectric properties of the phosphorene-WSe$_2$ under hydrostatic pressure.

This paper is organized as follows. In Sec.~\ref{Heterostructure} we present the geometry of the phosphorene-WSe$_2$ heterostructure, introduce the notion of the vertical pressure applied to the heterostructure, and provide all the relevant numerical details for the performed first-principles calculations. In Sec.~\ref{Bandstructure} the effect of applied pressure on electronic bands of the phosphorene-WSe$_2$ heterostructure is discussed with a particular emphasis on the spin-orbit coupling, spin relaxation of phosphorene holes close to the $\Gamma$ point, and thermoelectric properties. Finally, Sec.~\ref{conclusions} summarizes the most important findings.

\section{Atomic structure and first principles calculation details}\label{Heterostructure}
\begin{figure}[t]
\centering
\includegraphics[width=0.49\textwidth]{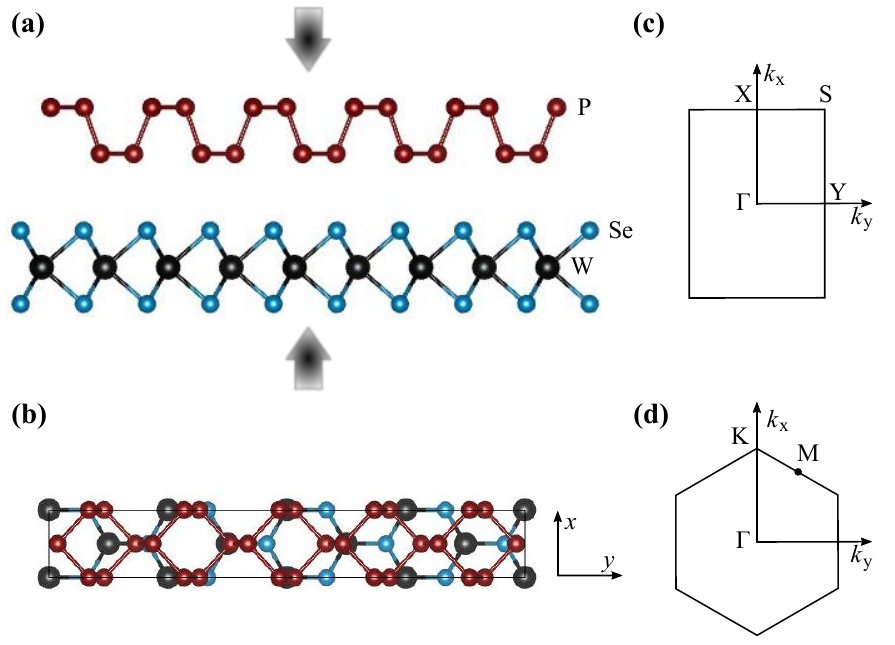}
\caption{Side~(a) and top~(b) view of the phosphorene-WSe$_2$ heterostructure subjected to the vertical pressure. The heterostructure is constructed in such a way that its $x$/$y$ direction corresponds to the zigzag/armchair edge of the phosphorene ML. In~(c) and~(d), the Brillouin zones of phosphorene and WSe$_2$ MLs, respectively,  are shown, with marked high symmetry points.}
\label{FigHet}
\end{figure}

In Fig.~\ref{FigHet}~(a)-(b), we present top and side views of the atomic structure models of the phosphorene-WSe$_2$ heterostructure subjected to hydrostatic pressure $P$. The supercell of the considered heterostructure contains 20 phosphorene (5 unit cells), 8 tungsten, and 16 selenium atoms and was constructed using the CellMatch code~\cite{L15}. 
Since we focus on the properties of phosphorene, we keep its lattice vectors unstrained. A minor strain of  0.51\% was applied to WSe$_2$ to satisfy the commensurability condition required by the periodic simulation cell. Lattice vectors of the phosphorene monolayer (ML) are equal to ${\bm a}=a~{\bf e}_x$, ${\bf b}=b~{\bf e}_y$,
where $a=3.2986~\AA$ and $b=4.6201~\AA$~\cite{JKG+19}, while the lattice vectors of the WSe$_2$ ML correspond to ${\bm a}_1=a_{\rm W} {\bf e}_x$,  ${\bm a}_2=a_{\rm W}(-{\bf e}_x+\sqrt{3}{\bf e}_y$)/2 ($a_{{\rm W}}=3.286\AA$~\cite{WJ69}). The corresponding Brillouin zones (BZs) of the heterostructure constituents are depicted in Fig.~\ref{FigHet}~(c) and~(d).

First-principles calculations of the phosphorene-WSe$_2$ heterostructure were performed using the plane wave Q{\sc{uantum}} ESPRESSO (QE) package~\cite{QE1,QE2}. The Perdew–Burke–Ernzerhof exchange-correlation functional was employed~\cite{PBE96}. Atomic relaxation was performed using the quasi-Newton scheme and scalar-relativistic SG15 optimized norm-conserving Vanderbilt (ONCV) pseudopotentials~\cite{HSC79,H13,SG15,SGH+16}. For ionic minimization, the force and energy convergence thresholds were set to $1\times10^{-4}$~Ry/bohr and $10^{-7}$ Ry, respectively. Additionally, the Monkhorst-Pack scheme with a $56\times 8$ $k$-point mesh was used, small Methfessel-Paxton energy level smearing of 1~mRy~\cite{MP89}, and kinetic energy cut-offs for the wave function and charge density 80\,Ry and 320\,Ry, respectively. We performed our calculations using three different van der Waals corrections: Grimme's DFT-D2 (D2)~\cite{G06,BCF+09}, non-local rvv10~\cite{VV10,SGG13}, and Tkachenko-Scheffler (TS)~\cite{TS09}. 
In all cases, a vacuum of 20\,${\AA}$ in the $z$-direction was employed. For the relaxed structures, the average distance between the bottom phosphorene and the closest selenium plane in the $z$-direction is equal to $3.31\AA$, $3.41\AA$, and $3.66\AA$ for the vdW corrections D2, rvv10, and TS, respectively.
Application of the hydrostatic pressure affects mostly the interlayer distance between the phosphorene and WSe$_2$. The calculation procedure we adopted is as follows: 
we started from the equilibrium structure and reduced the vertical distance between phosphorene and WSe$_2$ monolayer by $d_{\rm p}$. Next, we relaxed both the lattice parameters and atomic positions while fixing the $z$-components of the outermost phosphorene and selenium atoms. In this way, the relaxation of lateral lattice parameters lowers the in-plane stress in the heterostructure. Constraining the thickness by fixing the outermost atoms, a pressure with a dominant vertical component arises.

In the case of noncollinear density functional theory (DFT) calculations with spin-orbit coupling, the $k$-point mesh and kinetic energy cutoffs for the wave function and charge density were the same as in the relaxation calculations. We used fully relativistic SG15 ONCV pseudopotentials and increased the energy convergence threshold to $10^{-8}$~Ry. In all cases, the dipole correction~\cite{B99} was taken into account to properly determine the energy offset due to dipole electric field effects. Finally, we have performed a series of noncolinear DFT calculations with a nonzero electric field applied perpendicular to the heterostructure plane. The presence of such a field was modeled by a zig-zag potential that changes its slope in the vacuum region, with strengths from $-1~{\rm V/nm}$ to $1~{\rm V/nm}$~\cite{B99}.
\begin{table}[t]
\caption{The modifications of the vertical pressure and the percentage changes of the in-plane dimensions of the heterostructure cell in the $x$ and $y$ direction, $\delta_x$  and $\delta_y$, upon reducing the vertical distance $d_{\rm p}$. We performed relaxation calculations using three different vdW corrections:  Grimme-D2,  non-local rvv10, and Tkachenko-Scheffler.}\label{TAB:pressure}
\centering
\small
\setlength{\tabcolsep}{7pt}
\renewcommand{\arraystretch}{1.0}
\begin{tabular}{ccccc}
\hline\hline
vdW & $_{\rm p}$ [\AA] & $P$ [kbar]& $\delta_x$ [\%] & $\delta_y$ [\%] \\\hline
\multirow{6.1}{*}{D2}&0.1 &2.59&0.71 & -0.42 \\
&0.2 &5.57&0.77 & -0.31 \\
&0.3 &9.02&0.86 & -0.17 \\
&0.4 &12.59&0.95& -0.02 \\
&0.5 &16.50&1.04&  0.18 \\\hline
\multirow{6.1}{*}{rvv10}&0.1 &2.04 &0.96&  -0.19 \\
&0.2 &4.69 &1.02&  -0.10 \\
&0.3 &7.94 &1.10&  -0.02 \\
&0.4 &11.55 &1.18& 0.17 \\
&0.5 &15.68 &1.28&  0.36 \\\hline
\multirow{6.1}{*}{TS}&0.1 &1.28 &0.67&  -0.72 \\
&0.2 &2.56 &0.70&  -0.69 \\
&0.3 &4.07 &0.73&  -0.65 \\
&0.4 &4.95 &0.74&  -0.66 \\
&0.5 &7.99 &0.84&  -0.52 \\\hline\hline
\end{tabular}
\end{table}
\begin{figure*}[t]
\includegraphics[width=0.98\textwidth]{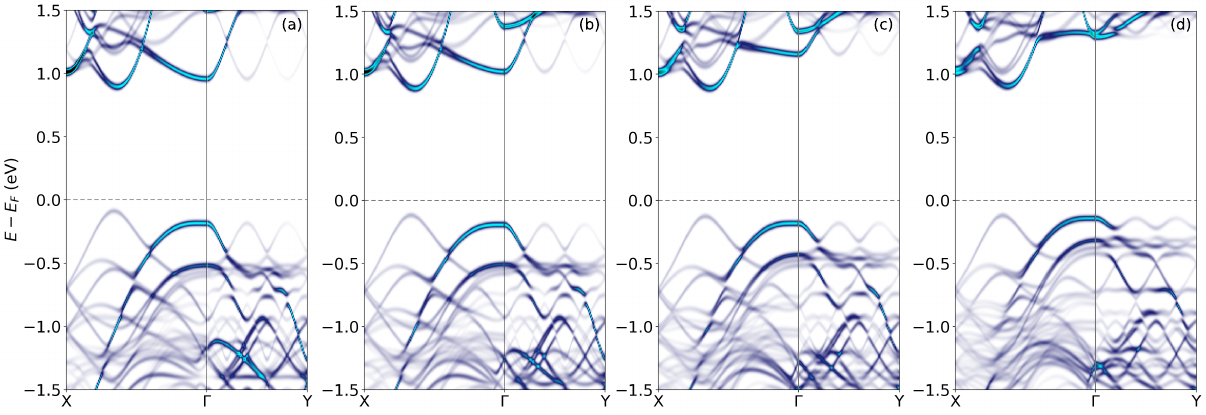}
\caption{Comparison of the hydrostatic pressure influence on the band structure in the phosphorene-WSe$_2$ heterostructure with the fully relaxed one~(a), unfolded to the X$\Gamma$Y path of the phosphorene Brillouin zone.  In panels (b)-(d), the phosphorene-WSe$_2$ heterostructure was exposed under the hydrostatic pressure equal to 2.59 kbar, 9.02 kbar, and 16.50 kbar, respectively. 
In all cases, the heterostructures were relaxed using the QE code and the Grimme-D2 vdW correction.
}\label{FigUnfold}
\end{figure*}

\section{Analysis of the pressure modulated band structure}\label{Bandstructure}
In Table~\ref{TAB:pressure} we present the calculated pressure values for $d_{\rm p}$ ranging from 0.1 $\AA$ to 0.5 $\AA$.
Since lateral lattice parameters are free to change,
we also present the relative changes $\delta_x$ and $\delta_y$ of the in-plane dimensions of the heterostructure cell in the $x$ and $y$ directions. Note that for the minimal unit of phosphorene, the changes in the $x$ and $y$ direction correspond to $\delta_x$ and $\delta_y/5$, respectively.
The pressure increases roughly linearly when reducing the heterostructure's effective thickness, with values in the range of the experimental setups (18 kbar in the graphene-WSe$_2$ heterostructure~\cite{FMZ+21}).

The discrepancies in the calculated pressure using different vdW corrections come solely from the unequal initial interlayer distances of the relaxed heterostructure using D2, rvv10, and TS vdW corrections. Thus, although the absolute change of the heterostructure thickness is the same for all three vdW corrections, the relative change is not the same. That is why the calculated pressure in the heterostructure relaxed using the TS vdW correction, having the sizable larger interlayer distance when compared to the D2 and rvv10 cases, is much lower than in the other two cases. 

To analyze the changes induced by pressure on the electronic properties of the phosphorene-WSe$_2$ heterostructure, in Fig.~\ref{FigUnfold}, we present the band structure unfolded to the XGY path of the Brillouin zone of phosphorene in four distinct cases (we assume D2 vdW correction in all 4 cases): (a) heterostructure without pressure, (b)-(d) heterostructure subjected to the pressure of 2.59, 9.02, and 16.50 kbar, respectively. 
The most prominent effect of the applied pressure is the increase of the direct band gap at the $\Gamma$ point, which suggests a weakening of the interaction between the valence and conductance bands. The top valence phosphorene band dispersion around the $\Gamma$ point remains almost unchanged. The K valley band of WSe$_2$ maps along the $\Gamma$X direction, and the $\Gamma$ valley band forms the lower-lying valence band. Under pressure, the lower-lying valence band moves closer to the top valence phosphorene band. Therefore, one can expect an increased spin-orbit coupling proximity effect in phosphorene holes. Hybridization of the K valley WSe$_2$ band and phosphorene hole band at about $-0.4$~eV is responsible for transferring the $z$-component of spin to the phosphorene.

To get a quantitative insight into the changes in the spin-orbit proximity effect due to the applied pressure, in what follows we will analyze the spin-orbit field of phosphorene holes around the $\Gamma$ point and compare it to the values obtained for the fully relaxed heterostructure configuration~\cite{MGK+23}.

\subsection{Effective parameters}\label{effectivepar}
The spin physics of phosphorene holes around the $\Gamma$ point can be well described in terms of the effective spin-orbit coupling Hamiltonian compatible with the ${\bf C}_{1{\rm v}}$ symmetry of the heterostructure~\cite{MGK+23,MGK++23},
\begin{equation}\label{effectiveHam}
 H_{\rm eff}=\lambda_1 k_x \sigma_y+\lambda_2 k_y \sigma_x + \lambda_3 k_x \sigma_z.
\end{equation}
The Hamiltonian~\eqref{effectiveHam} includes terms linear in momenta, as well as the Pauli spin operators $\sigma_i$, $i=x,y,z$, which are connected to the spin operators $S_i$ through the relation $S_i=\hbar/2 \sigma_i$. Whereas the $\lambda_1 k_x \sigma_y+\lambda_2 k_y \sigma_x$
part of the total spin-orbit field corresponds to the in-plane spin components and can be created simply by applying the perpendicular electric field, the out-of-plane component $\lambda_3 k_x \sigma_z$ is a signature of the mixed-lattice heterostructure that triggers the effective in-plane electric field into the phosphorene monolayer.

Although the effective form of the spin-orbit coupling Hamiltonian can be deduced from symmetry, the spin-orbit coupling parameters $\lambda_1$, $\lambda_2$, and $\lambda_3$ need to be determined by fitting the model~\eqref{effectiveHam} to the DFT data. This was done in the following way: for each studied configuration, the DFT data about the spin splitting energy and spin expectation values of the top valence band around the $\Gamma$ point were fitted to the spin-orbit coupling Hamiltonian model. For the fitting, we considered the X$\Gamma$ and $\Gamma$Y paths, up to the distance $0.009\,\AA^{-1}$ away from the $\Gamma$ point. 

\begin{figure}
    \centering
    \includegraphics[width=0.99\columnwidth]{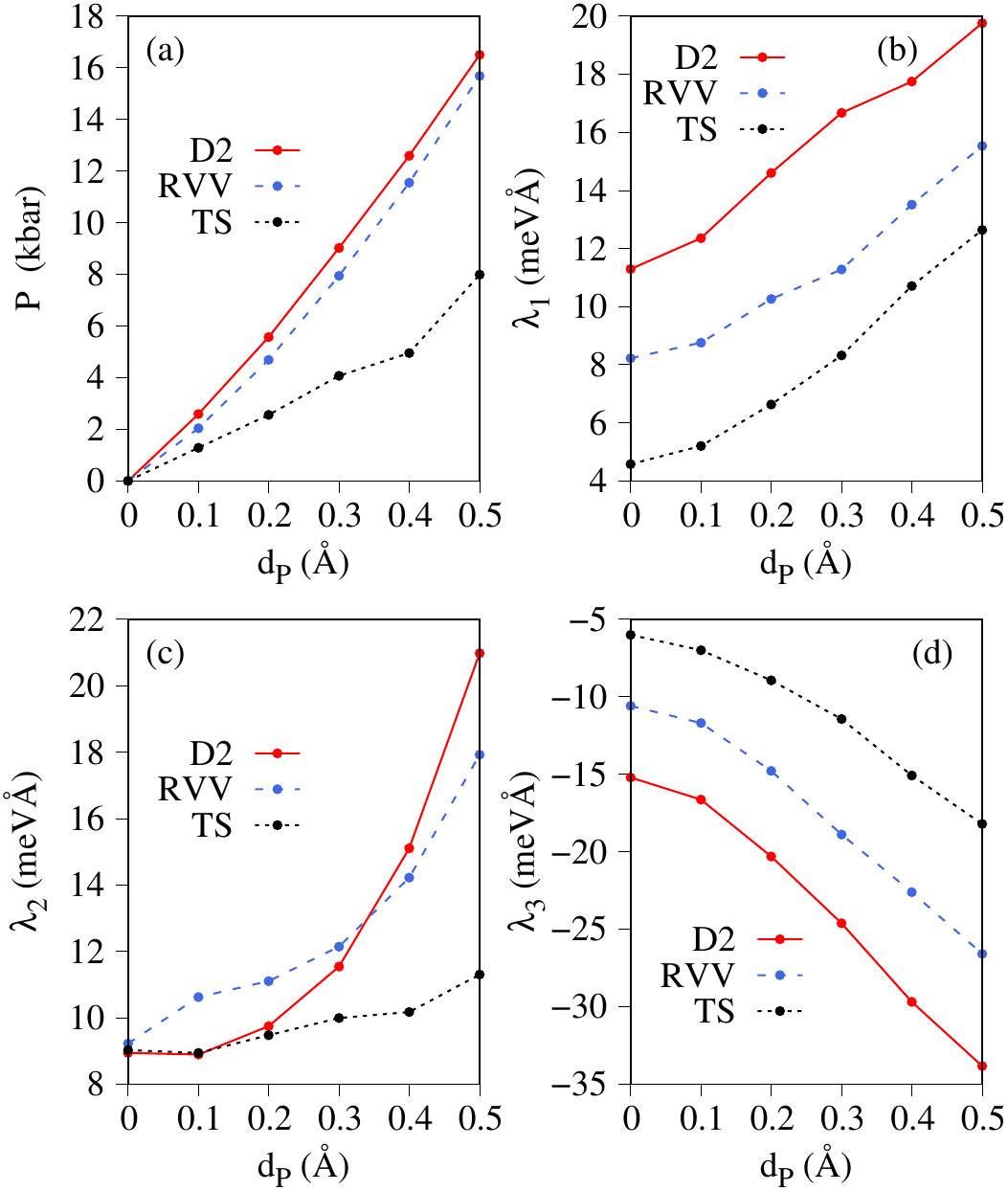}
    \caption{Dependence of the pressure (a) and spin-orbit coupling parameters of phosphorene holes~(b-d) around the $\Gamma$ point on the interlayer distance change $d_{\rm p}$ for three considered vdW corrections.}
    \label{fig:lambda_vs_dp}
\end{figure}
The obtained spin-orbit coupling parameters as a function of the $d_{\rm p}$ for phosphorene holes are given Fig.~\ref{fig:lambda_vs_dp}(b-d). The pressure dependence on $d_{\rm p}$, given in Table~\ref{TAB:pressure}, is also plotted in Fig.~\ref{fig:lambda_vs_dp}(a).
Comparing the spin-orbit coupling parameters values to the zero pressure case ($d_{\rm p}=0$) \cite{MGK+23} one sees that the pressure enhances the spin-orbit coupling $\lambda$ parameters by a factor of three. This is expected since the acquired spin-orbit coupling of phosphorene holes comes due to the proximity effect, which depends on the interlayer distance~\cite{MGK+23}. 
Thus, the presence of vertical pressure decreases the distance between the constituent MLs (strengthens the interlayer hopping) and at the same time increases the proximity-induced spin-orbit coupling. The controllability of the distance between the MLs through the vertical pressure suggests that pressure can be used as an experimentally accessible fine-tuning knob for the control of the spin physics in proximitized phosphorene. 
Although demonstrated in the example of phosphorene-WSe$_2$ heterostructure with zero twist angle, the obtained results can be easily extended to the 60-degree twist angle case. As shown in~\cite{MGK+23}, the dominant difference in the spin texture of phosphorene holes around the $\Gamma$ point in the case of zero and 60-degree twist angle lies in the sign change of the $\lambda_3$ parameter only due to the discovered~\cite{MGK+23} valley-Zeeman nature of the spin-orbit field $k_x \sigma_z$. Our results also suggest that the vertical pressure does not lead to a qualitative change in the type of spin texture induced due to the proximity effect. Thus, the tunable increase of the spin-orbit proximity effect can be expected in phosphorene-based trilayer heterostructures, in which phosphorene is symmetrically or asymmetrically encapsulated by two WSe$_2$ monolayers~\cite{MGK++23}.

\begin{figure}
    \centering    \includegraphics[width=0.8\columnwidth]{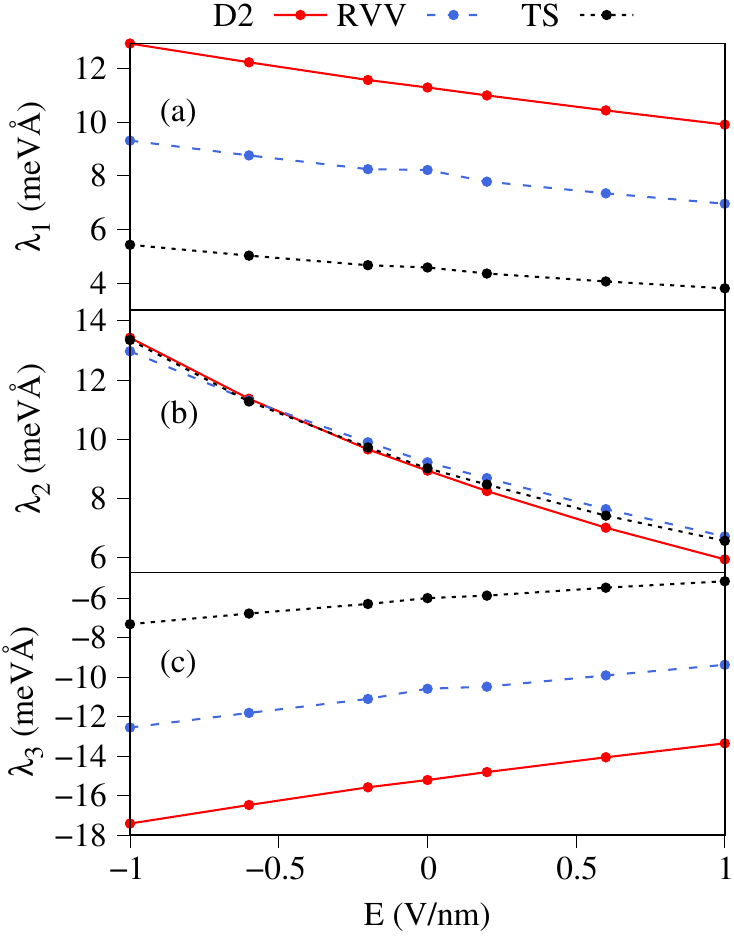}
    \caption{Dependence of spin-orbit parameters of the model Hamiltonian (\ref{effectiveHam}) on the transverse electric field amplitudes for three considered vdW correction types.}
    \label{fig:lambda_vs_E}
\end{figure}

Finally, we discuss and compare the fine-tuning of the spin-orbit proximity effect using hydrostatic pressure with the electric field tuning of the spin-orbit coupling. As known from Ref.~\cite{KGF16}, the Rashba effect in phosphorene leads to the appearance of the in-plane spin texture only, triggered by the broken inversion symmetry. Besides that, in the phosphorene-WSe$_2$ heterostructure, there is an additional possibility to tune the out-of-plane spin-orbit field with the electric field.
By performing a series of calculations for the relaxed heterostructure using the D2 vdW correction and the electric field strengths from -1 V/nm to 1 V/nm in steps of 0.4 V/nm, we have determined the $\lambda$ parameters in all the studied cases, obtained after fitting the spin-orbit Hamiltonian model~\eqref{effectiveHam} to the DFT data; see Fig.~\ref{fig:lambda_vs_E}. The calculations show that the negative electric field increases all the components of the spin-orbit field, although not as efficiently as when the vertical pressure is applied. 
Thus, pressure represents a promising alternative to the traditional approaches of modulating the spin-orbit coupling strength in proximitized materials, such as the electric field tuning utilizing the Rashba effect.
\begin{figure*}[t]
    \centering    
    \includegraphics[width=0.84\textwidth]{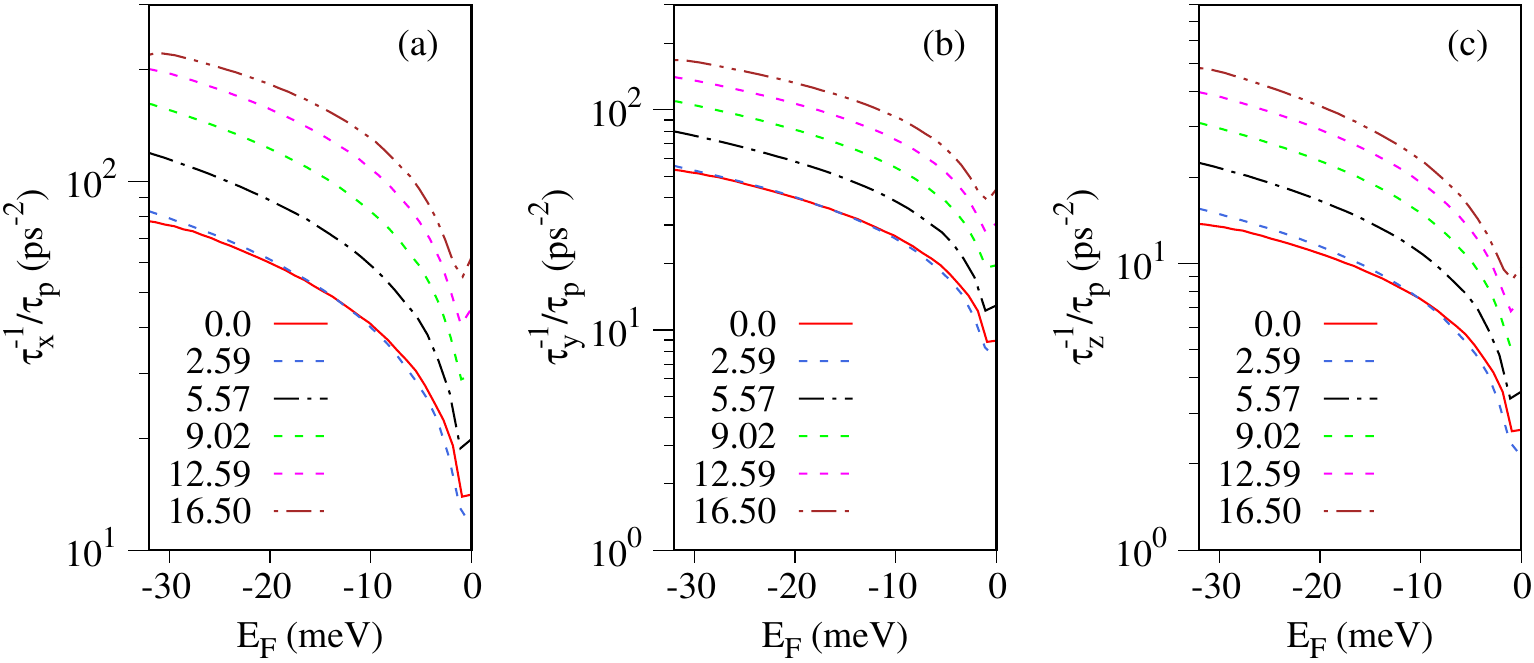}
    \caption{Calculated average spin relaxation rate $\tau_{i}^{-1}$ of phosphorene hole states for spin components $i=\lbrace x, y, z\rbrace$,  in units of momentum relaxation time $\tau_p$ (ps), versus the position of the Fermi level $E_F$, measured from the top of the valence band, and for different values of hydrostatic pressure, given in kbar units. In all studied cases, the phosphorene-WSe$_2$ heterostructure was relaxed using the Grimme-D2 vdW correction.}
    \label{fig:DP_time}
\end{figure*}
\subsection{Pressure-modulated spin relaxation time of phosphorene holes}\label{spinrelaxation}

Spin relaxation times represent an experimentally measurable quantity that is dependent on the spin-orbit coupling strength. While in pristine phosphorene
the upper limit of spin relaxation times~\cite{ATK+17,CLT+24} comes from the Elliot-Yaffet mechanism~\cite{EY1954},
the proximity-induced spin-orbit coupling in phosphorene on a WSe$_2$ monolayer triggers the Dyakonov-Perel (DP) spin relaxation mechanism~\cite{DP1971}. Within the DP regime, the spin relaxation time $\tau_{i}$ can be calculated using the relation
\begin{equation}\label{spinrelformula}
\tau_{i}^{-1}=\Omega_{\perp,i}^2\tau_p,
\end{equation}
where $\Omega_{\perp,i}^2 = \langle \Omega^2\rangle - \langle \Omega^2_i\rangle$ corresponds to the Fermi surface average of the squared spin-orbit field component $\Omega_{{\bf k}\perp,i}^2$ that is perpendicular to the spin orientation $i=\{x,y,z\}$, while $\tau_p$ is the momentum relaxation time. From the first-principle calculations, one can directly extract the spin-orbit field $ \Omega_{\bf{k},i}$ for the spin-split top valence band of phosphorene at the given ${\bf k}$ point of the BZ using the relation~\cite{KF21}
\begin{equation}
    \Omega_{\mathbf{k},i}=\frac{\Delta_{\rm so}}{\hbar}\frac{s_i}{|s|},
\end{equation}
in which the $\Delta_{\rm so}$ corresponds to the spin splitting value, while $s_i$ is equal to the expectation value of the spin one-half operator at ${\bf k}$. Using the Fermi contour averaging formula
\begin{equation}
    \langle \Omega_i^2 \rangle = \frac{1}{\rho(E_F)S_{\rm BZ}}\int_{\rm FC}\frac{\Omega^2_{{\bf k},i}}{\hbar|v_F({\bf k})|},
\end{equation}
in which $S_{BZ}$ represent the Fermi surface area, $\rho(E_F)$ corresponds to the density of states per spin at the Fermi level, $v_F({\bf k})$ is the Fermi velocity, while the integration goes over an isoenergy contour, one can calculate  
$\Omega_{\perp,i}^2$ for different hydrostatic pressure values, and correspondingly, analyze the influence of the hydrostatic pressure on DP spin relaxation. 
Results of such calculations for the top valence band and three spin components are shown in Fig.~\ref{fig:DP_time}. 
First, we notice that the spin relaxation times in all three directions are on the same time scale, which can be rationalized by the same energy scale of the $\lambda$ parameters given in Fig.~\ref{fig:lambda_vs_dp}. 

For weak hole doping, when crystal momenta are relatively close to the $\Gamma$ point, the ratio $\Omega_{\perp,i}^2(d_{\rm p}=0.5~\AA)/\Omega_{\perp,i}^2(d_{\rm p}=0~\AA)$ is roughly 4 for all spin directions.  It  can be connected to factor 2 increase of $\lambda$  parameters with pressure, since $\Omega_{{\bf k},x/y/z}^2\sim s_{x/y/z}^2\sim (\lambda_{2/1/3})^2$. Increasing the doping to $E_{\rm F}\approx 30$ meV,  the ratio $\Omega_{\perp,i}^2(d_{\rm p}=0.5~\AA)/\Omega_{\perp,i}^2(d_{\rm p}=0~\AA)$  reduces to approximately 3. 
One could explain this modification with the increased influence of the higher in $k$ order corrections to the spin-orbit field when moving from the $\Gamma$ point, which is not included in~\eqref{effectiveHam}. 
Thus, we have established a clear correspondence between the increase in the spin-orbit coupling strength and the decrease of $\tau_i$. We believe that this correspondence can be exploited as a simple tool for the detection of the hydrostatic pressure-induced effects on spin-orbit coupling. 

\subsection{Influence of the lateral shift on the spin-orbit proximity effect}

As a next step, we investigate the influence of the lateral shift between the phosphorene and WSe$_2$ MLs on the proximity-induced spin-orbit effect in phosphorene. To compare the result with the case of the zero shift, we will analyze the lateral change in the $y$ direction that is compatible with the ${\bf C}_{1{\rm v}}$ symmetry of the heterostructure. Thus, we can investigate the lateral shift effect by comparing the changes in $\lambda$ parameters of phosphorene holes around the $\Gamma$ point. Concretely, we have analyzed the lateral shifts ${\bf r}_s$ from 0~$\AA$ to 2.4~$\AA$ in steps of 0.4 \AA, where 2.4~$\AA$ roughly corresponds to half of phosphorene's lattice parameter $b$ in the $y$-direction. Furthermore, in all the calculations, D2 vdW correction was assumed.
To investigate the combined role of the hydrostatic pressure and the lateral shift, besides the case of zero pressure, we have analyzed the cases of three different interlayer distance changes $d_{\rm p}$, 0.1 $\AA$,  0.3 $\AA$, and  0.5 $\AA$.
\begin{figure}
    \centering \includegraphics[width=0.9\columnwidth]{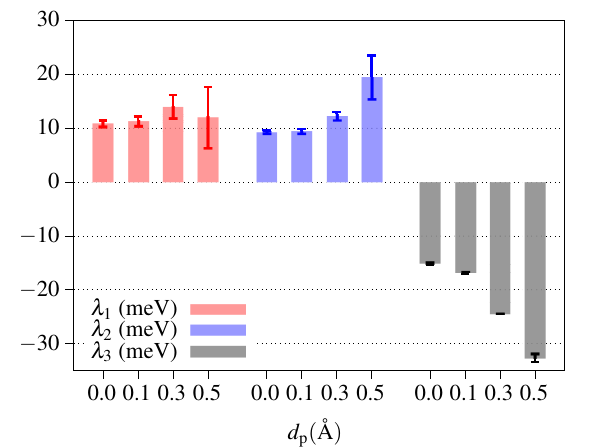}
    \caption{Influence of the lateral shift in the $y-$direction on the spin-orbit coupling parameters $\lambda_{1/2/3}$.
    The height of the bars corresponds to the mean value of the parameter amplitude while the error bars are mean square deviations calculated for a set of mutual lateral shifts of phosphorene with respect to WSe$_2$. The phosphorene-WSe$_2$ heterostructure was relaxed using the Grimme-D2 vdW correction in all studied cases.
    }
    \label{fig:lateralshift}
\end{figure}

The changes of spin-orbit parameters due to the lateral shift can be well understood using the average spin-orbit coupling strength $\lambda_i^{\rm aver}=\langle\lambda_i({\bf r}_s)\rangle$, $i=1,2,3$, and its error bar $\Delta \lambda_i$, defined as the mean square deviation $\Delta \lambda_i=\sqrt{\langle (\lambda_i^{\rm aver}-\lambda_i({\bf r}_s))^2\rangle}$ from the mean value $\lambda_i^{\rm aver}$. The results are gathered in FIG.~\ref{fig:lateralshift}, showing the negligible variation of spin-orbit parameters for zero and weak pressures, whereas for significantly strong vertical pressure strength there is a notable variance of $\lambda$ parameters. The nontrivial correlation between the hydrostatic pressure and the relative shift between phosphorene and WSe$_2$ monolayer can be explained by much stronger repulsion between the bottom phosphorene and top Se atoms when their relative distance is minimized due to the lateral shift.

\subsection{Thermoelectric effects}
Furthermore, we show that not only spin but also thermoelectric properties of the phosphorene-WSe$_2$ heterostructure, being dependent on the electronic structure of the studied system, can be successfully tuned with the hydrostatic pressure.
\begin{figure}[b]
    \centering 
    \includegraphics[width=1\columnwidth]{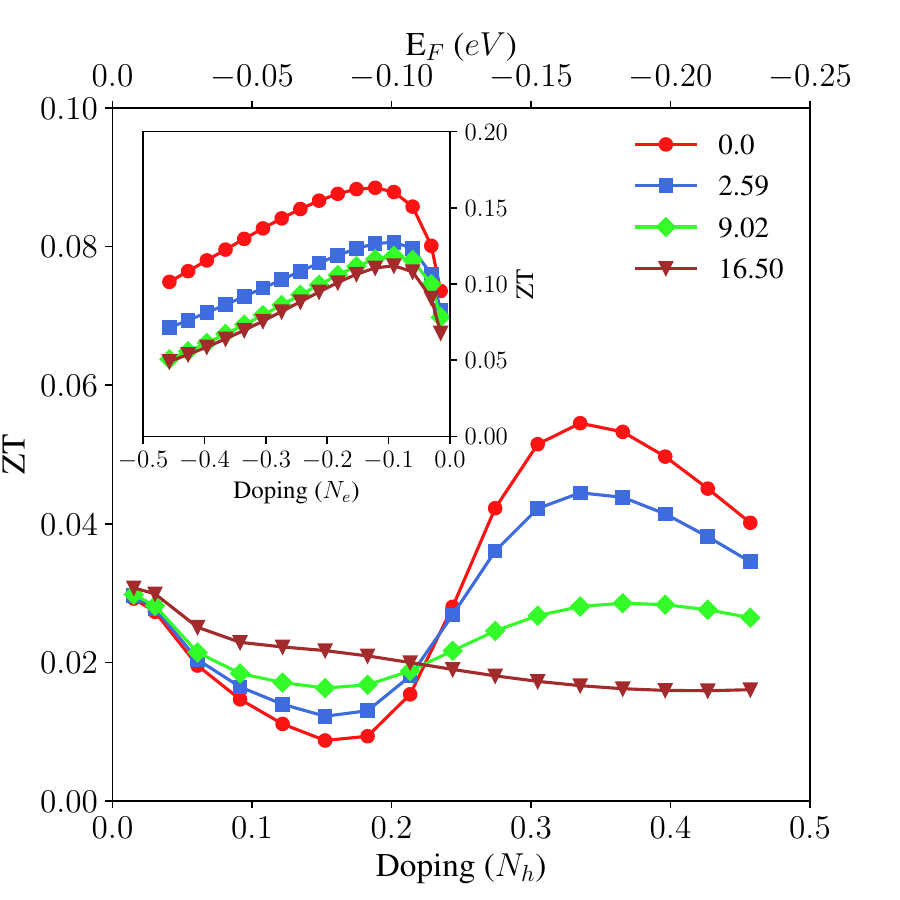}
    \caption{Calculated figure of merit ($ZT$) as a function of hole doping for various hydrostatic pressures given in kbar units at room temperature. The inset shows $ZT$ values for electron doping. The zero point of the Fermi level is aligned with the top of the valence band for hole doping and the bottom of the conduction band for electron doping. The carrier concentration ranges from $2 \times 10^{12}$ to $6 \times 10^{13}~{\rm cm^{-2}}$, corresponding to the minimum and maximum doping values shown in the figure for both electrons and holes.
    }
    \label{fig:ZT-v1}
\end{figure}
In Fig. \ref{fig:ZT-v1}, we present the figure of merit, $ZT$, at 300~K for phosphorene-WSe$_2$ heterostructure as a function of hole doping under various hydrostatic pressures.
The dimensionless figure of merit, $ZT$, characterizes the thermoelectric efficiency and is defined by the equation $ZT = S^2 \sigma T/(\kappa_e + \kappa_l)$, where $\sigma$ is the electrical conductivity, $S$ is the Seebeck coefficient or thermopower, $T$ is the absolute temperature, $\kappa_e$ is the electronic contribution to the thermal conductivity, and $\kappa_l$ is the lattice contribution to the thermal conductivity. The higher $ZT$ represents the higher efficiency of energy conversion. The Seebeck coefficient, electrical conductivity, and electronic thermal conductivity were calculated using Boltzmann transport theory as implemented in the BoltzTraP code \cite{Madsen2006:CPC}.

The phonon relaxation time for the flexural phonon mode ($z$-direction acoustic mode, ZA) for phosphorene has a minor contribution to the thermal conductivity attributed to the puckered hinge-like structure of phosphorene \cite{Qin2015:PCCP}. 
The stacking fault for the bulk WSe$_2$ reduces the thermal conductivity, affecting the group velocities, and limits the mean free path of the out-of-plane phonon modes \cite{Erhart2015:CM} via phonon localization and softening \cite{Chiritescu2007:Science}.
Detailed measurements of disordered bulk WSe$_2$ have reported a reduction of the thermal conductivity on the order of $1~{\rm W/(m\cdot K)}$ and interfacial thermal conductance $2.95~{\rm W/(mm^2\cdot K)}$ \cite{Easy2021:AMI}.
We note that the $\kappa_l$ ranges from 10 to $360~{\rm W/(m\cdot K)}$, depending on the specifics of the calculations \cite{Zhu2014:PRB,Xu2015:JAP,Jain2015:SR,Zhang2017:SR,Xu2018:MSMSE,Qin2018:Small,Lee2019:SR}.
Using micro-Raman spectroscopy the anisotropic thermal conductivities due to the anisotropic phonon dispersions were confirmed for few-layer black phosphorus and equal to $\sim 10~{\rm W/(m\cdot K)}$ and $\sim 20~{\rm W/(m\cdot K)}$ for zig-zag and armchair directions, respectively~\cite{Luo2015:NatComm}.
Therefore, we assume fixed value $\kappa_l = 15~{\rm W/(m\cdot K)}$ based on reported values.
For the momentum relaxation time, we use the experimental value $\tau_{\rm p} = 136~{\rm fs}$ \cite{ATK+17}.

As shown in Fig.~\ref{fig:ZT-v1}, $ZT$ initially decreases with increasing hole doping level up to about $0.2~e$. Since the temperature and $\kappa_l$ are fixed in our calculations, the only parameters affecting $ZT$ are the Seebeck coefficient, electrical conductivity, and electronic thermal conductivity. The electrical and electronic thermal conductivity increase with rising doping levels, but they do so at similar rates, such that their ratio remains constant (results not shown here). Therefore, the Seebeck coefficient is the primary factor influencing $ZT$ with doping.
It is accepted that the enhanced thermoelectric performance in 2D materials is mainly attributed to the increased Seebeck coefficient induced by the quantum confinement effect and reduced lattice thermal conductivity arising from strong phonon–phonon scattering~\cite{Duan2022:Small}.
In 2D materials, carriers are confined to a plane, leading to a Seebeck coefficient that differs from that in bulk materials. For 2D semiconductors, the Seebeck coefficient can be described by $S_{\rm 2D} = \frac{2 \pi^3 k_{\rm B}^2 T}{3eh^2} (\frac{m^*}{n})$, where $k_{\rm B}$, $h$, $n$, and $m^*$ are the Boltzmann constant, Planck constant, carrier concentration, and effective mass around the Fermi level, respectively \cite{Hicks1993:PRB,Snyder2008,Gao2020}. Here $S_{\rm 2D}$ is inversely proportional to $n$, highlighting a competition between $n$, and $m^*$. As doping increases, the carrier concentration rises, leading to a decrease in the Seebeck coefficient and, consequently, a reduction in $ZT$.
The turning point at a doping level of about $0.2~e$ is attributed to the contribution of phosphorene bands. As seen in Fig.~\ref{FigUnfold}, shifting the Fermi level to negative values initially intersects the WSe$_2$ bands. Further increases in energy toward the valence band incorporate contributions from phosphorene, which alters the Seebeck coefficient and results in the observed behavior in $ZT$. This effect can be explained by considering the difference in band curvature between phosphorene and WSe$_2$. 
Phosphorene has a flatter band structure compared to WSe$_2$, which implies a larger effective mass for charge carriers. The larger effective mass in phosphorene leads to a higher density of states near the Fermi level, which enhances the Seebeck coefficient. This increased Seebeck coefficient from phosphorene contributions compensates for the reductions from the WSe$_2$, leading to the observed turning point in the Seebeck coefficient and thus in $ZT$.
The calculated amplitudes of the $ZT$ for zero pressure are similar to the calculations for bare phosphorene \cite{Liao2015:PRB} suggesting a minor effect of the WSe$_2$.
Furthermore, increasing hydrostatic pressures from 0 to 16.50 kbar initially results in a less pronounced decrease in $ZT$ with increasing doping. This suggests that at lower doping levels, higher pressures can mitigate the reduction in thermoelectric efficiency. However, for doping levels beyond $0.2~e$, this trend reverses, and we observe that the structure under no pressure exhibits higher $ZT$ values compared to those under high pressure. This indicates that at higher doping levels, the beneficial effects of pressure on the materials' thermoelectric properties diminish, and pressure may even become detrimental to $ZT$.

A reverse trend is observed for electron doping, as shown in the inset of Fig.~\ref{fig:ZT-v1}. Here, $ZT$ initially increases with increasing doping levels due to the enhanced contribution of phosphorene bands (Fig.~\ref{FigUnfold}). This behavior also remains unchanged under different hydrostatic pressures, indicating that the thermoelectric properties of electron-doped systems are less sensitive to pressure changes.
However, under the highest hydrostatic pressures, $ZT$ is observed to be the lowest. This can be attributed to the upward shift of the phosphorene conduction bands caused by the applied pressure, which reduces the effective density of states available for thermoelectric transport, thereby lowering the $ZT$.
Moreover, it is noteworthy that the applied hydrostatic pressure does not affect the anisotropy of the Seebeck components, $S_{xx}/S_{yy}$, which is about 0.9 for electrons and 0.8 for holes at 300~K.
\begin{figure}[t]
    \centering 
    \includegraphics[width=1\columnwidth]{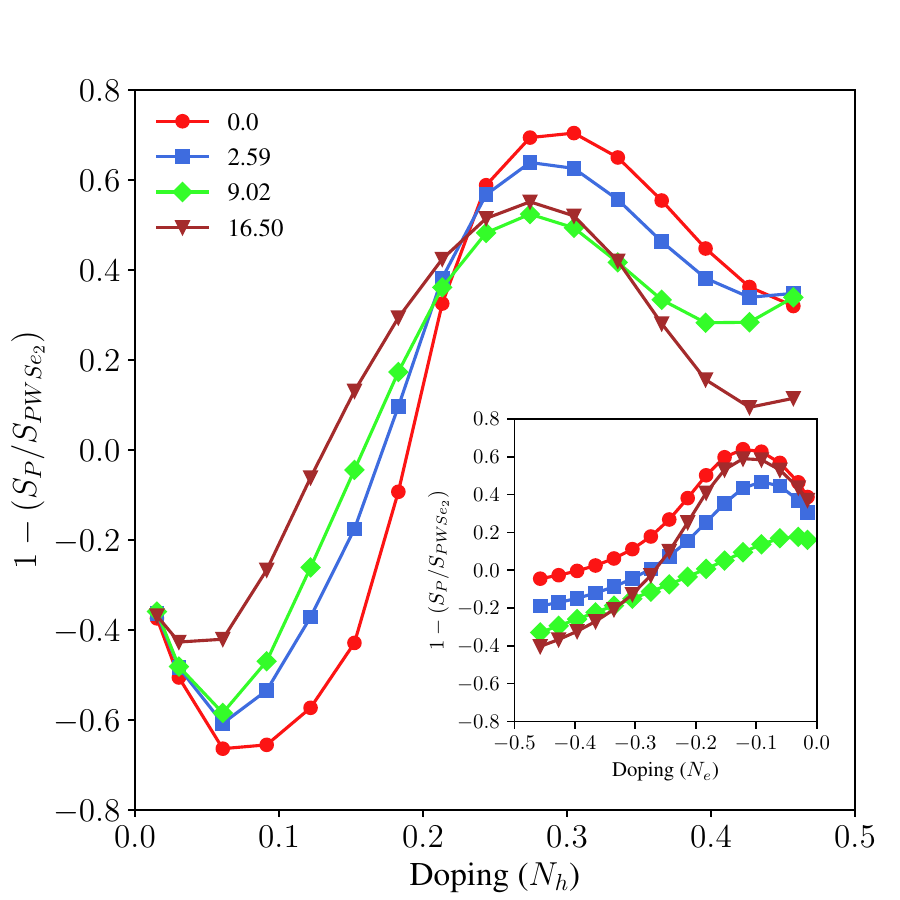}
    \caption{Calculated Seebeck coefficient relative change ($1 - S_{\rm P}/S_{\rm PWSe_2}$) as a function of hole doping for various hydrostatic pressures given in kbar units at room temperature. The inset shows the relative change for electron doping.
    }
    \label{fig:S-Ratio}
\end{figure}

The calculated relative change in Seebeck coefficients ($1 - S_{\rm P}/S_{\rm PWSe_2}$), where $S_{\rm P}$ is the Seebeck coefficient of the phosphorene layer alone (obtained by removing the WSe$_2$ layer) and $S_{\rm PWSe_2}$ is the Seebeck coefficient of the phosphorene-WSe$_2$ heterostructure, as a function of doping for various hydrostatic pressures at room temperature is shown in Fig.~\ref{fig:S-Ratio}.
At low hole doping levels (up to about $0.2~e$), the relative change is negative, suggesting that the Seebeck coefficient of the phosphorene layer alone is greater than that of the phosphorene-WSe$_2$ heterostructure.
Beyond $0.2~e$, the relative change becomes positive. This indicates that $S_{\rm P}$ is smaller than $S_{\rm PWSe_2}$ in this doping range, which aligns with the increased contribution from phosphorene bands to the overall thermoelectric performance of the heterostructure. The positive value suggests that the heterostructure's Seebeck coefficient is now dominated by the effects of the combined layers, with phosphorene playing a crucial role in enhancing the thermoelectric properties. The increase of $S_{\rm PWSe_2}$ with respect to $S_{\rm P}$ occurs exactly at the point where the phosphorene bands start contributing significantly to the heterostructure Seebeck coefficient, as shown in Fig.~\ref{fig:ZT-v1}. For electron doping the observed behavior of the Seebeck coefficient relative change also confirms the trends seen in the figure of merit $ZT$. Specifically, the enhancement of $ZT$ with increased electron doping is consistent with the positive value of the relative change at low doping levels.
\section{Conclusions}\label{conclusions}
We studied the influence of the vertical pressure in a mixed-lattice heterostructure made of phosphorene and a WSe$_2$ monolayer, which is a promising platform for the transfer of spin-orbit coupling from WSe$_2$, a strong spin-orbit coupling material, to phosphorene, via the proximity effect.
We focused on the spin physics of phosphorene holes around the $\Gamma$ point, for which we have 
extracted the parameters of the model spin-orbit Hamiltonian, compatible with the ${\bf C}_{1{\rm v}}$ symmetry of the phosphorene-WSe$_2$ heterostructure,
for several values of the hydrostatic pressure. 
We showed that for experimentally accessible pressure values, the increase in the spin-orbit coupling strength reaches a factor of two and is much more efficient than the prevalent electric-field tuning of the spin-orbit interaction. Finally, we have calculated the spin relaxation times of phosphorene holes related to the DP relaxation mechanism and showed that the gradual tuning of spin-orbit coupling strength by the hydrostatic pressure transfers to the tunable decrease of spin relaxation times, offering a simple route for indirectly determining the hydrostatic-pressure influence on the spin-orbit field in phosphorene using the spin relaxation times, an experimentally available quantity in heterostructures based on phosphorene.
Finally, our investigation of the thermoelectric properties reveals that the thermoelectric figure of merit, $ZT$, reduces with increasing pressure, and it is sensitive to hole and electron doping levels at 300~K. Specifically, $ZT$ decreases with increased doping, primarily due to changes in the Seebeck coefficient of phosphorene.

\acknowledgments
M.M. acknowledges the financial support
provided by the Ministry of Education, Science, and Technological Development of the Republic of Serbia. This project has received funding from the European Union's Horizon 2020 Research and Innovation Programme under the Programme SASPRO 2 COFUND Marie Sklodowska-Curie grant agreement No. 945478.
M.K.~acknowledges financial support provided by the National Center for Research and Development (NCBR) under the V4-Japan project BGapEng V4-JAPAN/2/46/BGapEng/2022. 
M.R. and M.G. acknowledge financial support provided by the Slovak Research and Development Agency under Contract No. APVV-SK-CZ-RD-21-0114.
M.G.~acknowledges financial support provided by Slovak Academy of Sciences project FLAG ERA JTC 2021 2DSOTECH and IMPULZ IM-2021-42.
D.L. acknowledges the support of CSF(GACR) project (24-11992S), and projects of the Ministry of Education, Youth and Sports No. LUASK22099, QM4ST CZ.02.01.01/00/22\_008/0004572, e-INFRA (ID:90254).
The authors gratefully acknowledge the Gauss Centre for Supercomputing e.V. 
for funding this project by providing computing time on the GCS Supercomputer SuperMUC-NG at Leibniz Supercomputing Centre.

\end{document}